\documentclass[reprint, aps, prb, amsmath, amssymb, floatfix]{revtex4-2}

\usepackage[utf8]{inputenc}
\usepackage{graphicx}
\usepackage{dcolumn}
\usepackage{bm}
\usepackage{booktabs}
\usepackage{siunitx}
\usepackage{xcolor}
\usepackage{placeins}
\usepackage{layouts}

\usepackage{hyperref}
\hypersetup{
    colorlinks=true,
    citecolor=blue,
    linkcolor=red,
    urlcolor=magenta
}

\usepackage{tikz}
\usetikzlibrary{positioning, shapes, arrows}

\usepackage[capitalise]{cleveref}
\crefname{equation}{Eq.}{Eqs.}
\crefrangeformat{equation}{Eqs.~(#3#1#4--#5#2#6)}

\renewcommand{\r}{\mathbf{r}}

\newcommand{\ads}{s}
\newcommand{\adx}{x}
\newcommand{\ady}{y}
\newcommand{\addelta}{\delta}
\newcommand{\adeta}{\eta}

\newcommand{\adtau}{\tau}
\newcommand{\f}{\mathbf{f}}

\newcommand{\ex}{\mathbf{e}_x}
\newcommand{\ey}{\mathbf{e}_y}
\newcommand{\ez}{\mathbf{e}_z}

\begin{document}

\title{Dynamic versus quasi-static response of \\ a cantilevered beam rotated harmonically} 

\author{Gilad Yakir} 
\author{Eduardo Gutierrez-Prieto} 
\author{Pedro M. Reis\footnote{To whom correspondance should be sent: pedro.reis@epfl.ch}} 

\affiliation{Flexible Structures Laboratory, Institute of Mechanical Engineering, École Polytechnique Fédérale de Lausanne (EPFL), 1015 Lausanne, Switzerland}

\begin{abstract}
We investigate a cantilevered elastic beam subjected to harmonic rotational motion. In the rotating frame, the beam experiences centrifugal and Euler fictitious forces, with negligible Coriolis effects. We validate a reduced-order \textit{elastica} model through precision experiments on slender beams rotating with a controlled sinusoidal angular velocity. Systematically exploring the parameter space, we identify regimes where inertial effects are negligible, enabling a quasi-static treatment despite harmonic driving. We characterize the transition to dynamic response using two dimensionless parameters, the Euler and centrifugal numbers, which compare centrifugal and Euler forces to bending forces. Counterintuitively, the quasi-static regime expands as rotational speed increases: faster rotation produces less dynamic response. The critical Euler number separating these regimes remains constant at low centrifugal numbers but follows square-root scaling at higher rotation rates, a transition driven by centrifugal stiffening. Our results establish the conditions under which quasi-static approximations remain valid for rotating flexible beams under harmonic driving.
\end{abstract}

\maketitle
\clearpage

\section{Introduction}
\label{sec: Introduction}

Nearly every machine built since the Industrial Revolution involves rotating components \cite{shigley1985mechanical}, including wheels, gears, fans, and turbines. These components have become so ubiquitous that they often go unnoticed. Regardless of their size and purpose, the dynamics of rotating systems is described by a set of differential equations in a frame of reference (FoR) \cite{meriam2020engineering}, which can be either inertial or non-inertial. Analyzed in a non-inertial frame, rotating systems are governed by three fictitious forces: \textit{centrifugal}, \textit{Euler}, and \textit{Coriolis} forces. The \textit{centrifugal force} scales with the square of angular velocity and acts radially outward. The \textit{Euler force} acts tangentially, opposite to angular acceleration, and arises from changes in angular velocity. The \textit{Coriolis force} acts perpendicular to both the angular velocity and the relative velocity. Although there is substantial literature on the structural response of rotating beams under steady rotation \cite{thomas2016hardening,apiwattanalunggarn2003finite,das2007free,lacarbonara2012geometrically,turhan2009nonlinear,da1986nonlinear,CILENTI2024104582}, relatively few studies have examined systems with unsteady angular velocities where Euler forces become important.

The effect of fictitious forces on point masses is well-established and commonly taught in introductory \textit{Dynamics} courses. However, their influence is more nuanced for rotating structures, where these forces interact with the body's geometric and material properties to dictate the mechanical response. Early foundational work on rotating beams and shafts was pioneered by Rankine and Jeffcott \cite{rankine1869centrifugal, jeffcott1919xxvii}; the latter's work, in particular, is often regarded as the first fundamental theory of rotordynamics.

Rotation critically affects the vibration characteristics of rotating beams \cite{yoo_vibration_1998,hoskoti2023modeling}; centrifugal forces cause stretching, which increases the bending stiffness and shifts natural frequencies and mode shapes \cite{yoo_vibration_1998,wright1982vibration,behzad2004effect}. When rotational acceleration is considered, the Euler force induces tangential deformation, introducing acceleration-dependent coupling in the governing beam equations \cite{kim_nonlinear_2016}. Recent work by \cite{clarabut_nonlinear_2023} further demonstrated these effects for a pitching cantilever beam subjected to both constant and time-varying angular velocities, showing that incorporating Euler forces significantly affects the beam's dynamic response. Notably, linearized geometric approximations become inadequate in this regime, failing to accurately predict both the bending deflection amplitude and the twist angle. This underscores the need to account for Euler forces in rotating structures.

We previously investigated rotating cantilevered and bistable beams under constant angular acceleration, characterized by linear velocity profiles, focusing on the resulting buckling behavior \cite{gutierrez-prieto_gyrophilia_2023}. For a cantilever beam clamped radially with the free end pointing to the center of rotation, the results revealed a critical interplay between the beam's natural curvature and the applied Euler force in dictating the buckling direction. For a bistable beam, on-demand snap-through transitions could be induced by Euler forces, enabling controlled switching between stable states. We later demonstrated that dynamic driving protocols enable independent control of bistable elements for programmable mechanical memory \cite{gutierrez-prieto_dynamic_2025}. Despite these contributions, significant knowledge gaps remain in understanding how velocity-dependent centrifugal forces interact with acceleration-dependent Euler forces under time-varying conditions. While prior studies have typically focused on steady or simplified rotational profiles, few have systematically explored the transition between dynamic and quasi-static responses in harmonically rotated cantilever beams or revealed how intrinsic mechanics suppress inertial effects under unsteady conditions.

Here, we assess the validity of a quasi-static approximation for unsteadily rotated cantilever beams. Quasi-static approximations offer significant computational advantages and are preferred in engineering analysis \cite{siddiqui2017quasi,ahmed2022parametric,burton2011wind,rajagopal2019quasi}. However, the validity conditions of this approximation for rotating structures remain largely unexplored. We adopt an existing reduced-order model based on Euler's \textit{elastica} in a rotating frame \cite{gutierrez-prieto_gyrophilia_2023} and validate it against precision experiments. We identify a regime where the beam's response can be accurately predicted by assuming quasi-static behavior under instantaneous forcing. These results highlight the interplay between centrifugal and Euler forces, establishing when simplified models are suitable for analyzing and designing rotating elastic systems. We then systematically explore material and geometric parameters, enabling us to perform a theoretical analysis that predicts the transition between quasi-static and dynamic behavior in slender rotating beams.

Our manuscript is structured as follows. \Cref{sec:Problem Definition} defines the system and outlines the research questions. \Cref{sec: Theoretical formulation,sec: methodology - numerical} revisits the reduced-order rotating \textit{elastica} model from \cite{gutierrez-prieto_gyrophilia_2023} and describes its implementation, while \cref{sec: methodology - experiments} presents the experimental apparatus and protocols. We validate the numerical model against experiments in \cref{sec: Validation of Numerics Against Experiments}, then define the quasi-static-to-dynamic boundary and identify where quasi-static assumptions are valid in \cref{sec:Defining the boundary between quasi-static and dynamic responses}. \Cref{sec: Scaling law of the phase boundaries} derives a theoretical description of the boundary, showing that centrifugal stiffening governs this transition. Finally, \cref{sec: conclusions} concludes and suggests future directions.
    
\section{Problem definition} \label{sec:Problem Definition}

\begin{figure}[h!]
    \centering
    \includegraphics[width=0.9\columnwidth]{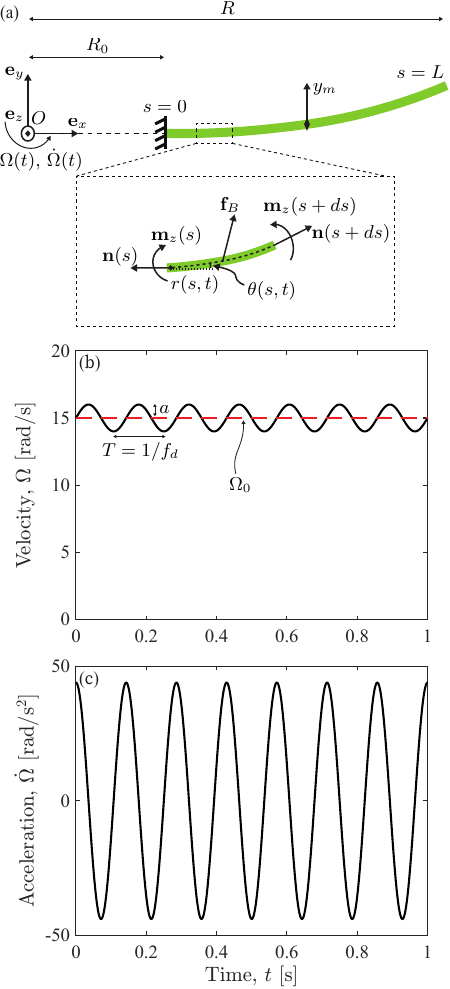}
    \caption{Schematic of a rotating cantilevered beam. (a) The slender beam  (arclength $L$) is clamped at a distance $R_o$ from the center of rotation, $O$, with the system rotating about $\ez$ at angular velocity $\Omega(t)$ and acceleration $\dot{\Omega}(t)$. Inset:  Infinitesimal beam element at arclength $s$ showing fictitious body force $\mathbf{f}_B=\f_c+\f_\omega+\f_e$; cf. Eqs.~ (\ref{eq:ForceC}-\ref{eq:ForceE}), internal tension $\mathbf{n}$ and moment  $\mathbf{m}_z$. (b) Representative sinusoidal angular velocity profile $\Omega(t)$ with mean $\Omega_0 = 15\,\text{rad}/\text{s}$, amplitude $a = 1\,\text{rad}/\text{s}$, and frequency $f_d = 7\,\text{Hz}$. (c) Corresponding angular acceleration $\dot\Omega$.} 
    \label{fig:Fig_1}
\end{figure}

We investigate the elastic deformation of a cantilevered beam mounted on an unsteadily rotating FoR, as shown schematically in \cref{fig:Fig_1}(a). 
Our system is similar to that of \cite{gutierrez-prieto_gyrophilia_2023}. 
A rotating triad $(\ex,\,\ey,\,\ez)$ is attached to the center of rotation $O$. The beam is clamped at a distance $R_o$ from $O$ with its undeformed centerline along $+\ex$. The FoR is rotated with angular velocity $\Omega(t)\,\left[\text{rad}/\text{s}\right]$ and angular acceleration $\dot{\Omega}(t)=\text{d}\Omega(t)/\text{d}t\,\left[\text{rad}/\text{s}^2\right]$ (Fig.~\ref{fig:Fig_1}b--c). The beam has arclength $L$, rectangular cross-section (thickness $h$, width $b$, cross-sectional area $A=bh$, and moment of inertia $I=bh^3/12$), Young's modulus $E$, and density $\rho$. We assume the beam is slender ($h = \lambda_1 L$, with $\lambda_1 \ll 1$), inextensible, and unshearable. The cross-section has width $b = \lambda_2 h$, with $\lambda_2 > 1$, and deformations are purely in-plane ($z=0$). Using standard \textit{elastica} kinematics, the centerline is parameterized by $s\in[0,\,L]$, from the clamp ($s=0$) to the free end ($s=L$), with position
$
    \mathbf{r}(s,t)=x(s,t)\ex+y(s,t)\ey 
$
and tangent angle $\theta(s,\,t)$ relative to $\ex$ defined by
$
    \partial\mathbf{r}/\partial s = \mathbf{r}_{,s}=\cos\theta\, \ex+\sin\theta\,\ey .
$
The Cartesian coordinates of the centerline at $s$ are
\begin{equation} \label{eq:x_s_nad_x_y}
\begin{aligned}
    &\adx (\ads) =  \int_0^{\ads} \cos\theta(\ads')d\ads'\quad \text{and} \\  
    &\ady (\ads) = \int_{0}^{\ads} \sin\theta(\ads')d\ads'.
\end{aligned}
\end{equation}

The rotating beam experiences three \textit{fictitious} body loads, the Coriolis, centrifugal, and Euler forces:
\begin{align}
    \f_c(t) &= -2\rho A \Omega \ez\times \dot{\r}, \label{eq:ForceC} \\
    \f_\omega(t) &= \rho A \Omega^2 \r, \label{eq:ForceW} \\
    \f_e(t) &= -\rho A \dot{\Omega} \ez \times \r. \label{eq:ForceE}
\end{align}
\cite{gutierrez-prieto_gyrophilia_2023} previously studied a rotating cantilevered beam under constant angular acceleration, oriented to apply a compressive centrifugal force. This induced a buckling instability whose direction could be tuned via the Euler force. Here, we reverse the beam's orientation, making the centrifugal force tensile, and impose a \textit{harmonic} rather than a constant angular velocity. Despite this dynamic loading, we demonstrate that, under certain conditions, the system behaves quasi-statically. Our goal is to characterize the boundary where the quasi-static approximation breaks down.

We impose sinusoidal angular velocity profiles,
\begin{equation} \label{eq: ang_vel}
        \Omega(t) = a\sin{\left(2 \pi f_d t\right)} + \Omega_0,
\end{equation}
with corresponding angular acceleration
\begin{equation}\label{eq: ang_acc}
        \dot{\Omega} (t) = 2 \pi a f_d\cos{\left(2 \pi f_d t\right)},
\end{equation}
where $a$ is the amplitude, $f_d$ the drive frequency, and $\Omega_0$ the mean angular velocity. Representative profiles are shown in \cref{fig:Fig_1}(b,\,c).

Due to the beam's size and velocity, the Coriolis force $\f_c$ is negligible compared to centrifugal and Euler forces. Following \cite{gutierrez-prieto_gyrophilia_2023}, the ratio  $|\f_c|/|\f_\omega| \sim |(\dot{\mathbf{r}})_r|/(\Omega R)$ indicates that $\f_c$ is significant only when the beam's velocity in the FoR is comparable to the tangential velocity of the imposed rotation, which does not occur for the parameters we explore.

With Coriolis forces neglected, the driving profiles in \cref{eq: ang_vel,eq: ang_acc} control the relative effects of centrifugal and Euler forces by tuning the mean angular velocity $\Omega_0$ relative to the angular acceleration amplitude, $\alpha=2 \pi a f_d$. Unlike constant acceleration, harmonic driving allows the beam to reach a steady-state oscillatory response. We fix the velocity amplitude at $a = 1\,\text{rad/s}$, so that varying the drive frequency $f_d$ independently controls the angular acceleration magnitude $\alpha = 2 \pi f_d$. This allows us to discern the relative influence of the Euler and centrifugal forces.

We apply the theoretical framework from \cite{gutierrez-prieto_gyrophilia_2023} to perform numerical simulations validated against precision experiments. We identify regions in parameter space where the beam's response is either \textit{dynamic} (when inertial effects are significant) or \textit{quasi-static} (when instantaneous static forces accurately predict the response). We will demonstrate that quasi-static conditions occur under large centrifugal loadings; counterintuitively, faster rotation leads to less dynamic response.

In summary, we aim to answer two key questions: (i) When can a harmonically rotating structure produce a quasi-static response? (ii) When do inertial effects become significant enough to shift the behavior from quasi-static to dynamic?

\FloatBarrier

\section{Theoretical formulation} \label{sec: Theoretical formulation}

We adopt the \textit{elastica}-based model developed by \cite{gutierrez-prieto_gyrophilia_2023} for a thin, linear-elastic cantilevered beam in a rotating frame. For completeness, we briefly review this model, derived from force and moment balance on an infinitesimal element of the beam (see schematic in the inset of \cref{fig:Fig_1}a). The dimensional equations are nondimensionalized using the total arclength $L$ for length, the characteristic bending force $EI/L^2$ for forces, and the bending frequency scale $f_s =(2\pi)^{-1} \sqrt{EI/\rho A L^4}$ for time. (Hereafter, overdots denote time derivatives $\dot{b}(t)=\partial b/\partial t$ for dimensional time and $\dot{\tilde{b}}(\tau)=\partial \tilde{b}/\partial \tau$ for dimensionless quantities.) The first natural frequency of the beam is 
\begin{equation}
f_n=\beta_1^2f_s=\frac{\beta_1^2}{2\pi} \sqrt{\frac{EI}{\rho A L^4}},
\label{eq:natural_freq}
\end{equation}
where $\beta_1=1.875$ is the modal constant for the first mode \cite{meirovitch1967analytical}. The resulting dimensionless equations of motion are
\begin{align}
        &\ex:~~ n_{x,s} + 2\mathcal{I}\dot{y} 
        + \mathcal{C}(1 - \delta + \delta x) + \label{eq:EOM1} \\\notag
        &\hspace*{10em}{}\delta \mathcal{E} y
        = \ddot{x} + \eta \dot{x},  \\
    &\ey:~~n_{y,s}- 2\mathcal{I}\dot{x} 
        + \mathcal{E}\left(1 - \delta +\delta x\right) + \label{eq:EOM2}\\ \notag
        &\hspace*{10em}{}\delta \mathcal{C} y= \ddot{y} + \eta \dot{y},  \\
    &\ez:~~\theta_{,ss}-n_x\sin{\theta} + n_y \cos{\theta} = 0\, , \label{eq:EOM3}
\end{align}
where we use the notation $\partial \langle \cdot \rangle / \partial s= \langle \cdot \rangle_{,s}$ and $\partial^2 \langle \cdot \rangle / \partial s^2= \langle \cdot \rangle_{,ss}$ for spatial partial derivatives. The dimensionless quantities are arclength $s\in[0,\,1]$, time $\tau$, horizontal $x$ and vertical $y$ displacements, internal tension  $\mathbf{n}=n_x\,\ex+n_y\,\ey$, geometric ratio $\delta = L/R$, and viscous damping coefficient $\eta=\gamma L^2/\sqrt{\rho A \,E I}$  (in Appendix \ref{app:B}, we determine $\gamma$, which accounts for internal dissipation effects). The dimensionless centrifugal, Euler, and inertial numbers in  \crefrange{eq:EOM1}{eq:EOM3} are
 \begin{equation}\label{eq:centrifugal_number}
    \mathcal{C}(\tau) = \frac{\rho A R L^3}{EI} \Omega^2(\tau),
\end{equation}
\begin{equation}\label{eq:euler_number}
    \mathcal{E}(\tau) =  \frac{\rho A  R L^3}{EI}f_s\dot{\Omega}(\tau),
\end{equation}
\begin{equation}\label{eq:inertial_number}
    \mathcal{I}(\tau) = \frac{\rho AL^4}{EI} f_s \Omega(\tau),
\end{equation}
which quantify the corresponding forces (Eqs.~\ref{eq:ForceC}-\ref{eq:ForceE}) relative to the characteristic bending forces per unit length, $EI/L^3$.

Assuming zero natural curvature, the boundary conditions for our cantilevered configuration (Fig.~\ref{fig:Fig_1}) are
\begin{equation} 
\label{eq: BC}
\begin{split}
    &~\,x(0,\tau) = 0,~~ ~\,y(0,\tau) = 0,
    \\
    &n_x(1,\tau) = 0,~~ n_y(1,\tau) = 0,
    \\
    &~\theta(0,\tau) = 0,~~\, \theta_{,s}(1,\tau) = 0.
\end{split}
\end{equation}

\section{Methodology: Numerical simulations} \label{sec: methodology - numerical}

While we use the same governing \crefrange{eq:EOM1}{eq:EOM3} as \cite{gutierrez-prieto_gyrophilia_2023}, several differences require adapting the discretization procedure: our reversed beam orientation produces tensile rather than compressive centrifugal loading (cf.~\cref{sec:Problem Definition}); we impose different boundary conditions (Eq.~\ref{eq: BC}); and we apply harmonic rather than constant driving (Eqs.~\ref{eq: ang_vel}-\ref{eq: ang_acc}). The modified spatial discretization is outlined below and detailed in Appendix \ref{app:A}, with key changes in the integration limits (Eq.~\ref{eq:x_s_nad_x_y}) and discretization matrices $\mathbf{L}$, $\mathbf{U}$, and $\mathbf{K}$.

Discretizing \crefrange{eq:EOM1}{eq:EOM3} yields $N{-}1$ coupled nonlinear ODEs for the tangential angle vector $\mathbf{\Theta} = \left[\theta_1,...,\theta_{N{-}1}\right]^\text{T}$:
\begin{equation} \label{eq:discretizedEq}
    \mathbf{M} \ddot{\mathbf{\Theta}} + \mathbf{C} \dot{\mathbf{\Theta}} + \mathbf{K}\mathbf{\Theta} = \mathbf{f},
\end{equation}
where $\mathbf{M}$, $\mathbf{C}$, and $\mathbf{K}$ are the mass, damping, and stiffness matrices, and $\mathbf{f}$ is the forcing vector. All these quantities are defined in Appendix \ref{app:A}. After a convergence test (see Appendix \ref{app:C}), we use $N{=}40$ nodes and solve \cref{eq:discretizedEq} with MATLAB's stiff ODE solver \texttt{ode23tb}. 

Quasi-static solutions are obtained by setting all time derivatives in \crefrange{eq:EOM1}{eq:EOM3} to zero, reducing the problem to instantaneous static equilibrium under time-varying forces.  We solve these equations with MATLAB's boundary value problem solver, \texttt{bvp4c}. We explore drive parameters in the ranges: $\mathcal{C}\in[0.03,\, 787]$ and $\mathcal{E}\in[0.2,\, 44]$.

\FloatBarrier

\section{Methodology: Experiments}
\label{sec: methodology - experiments}

\begin{figure*}[t!]
\centering
    \includegraphics[width=\textwidth]{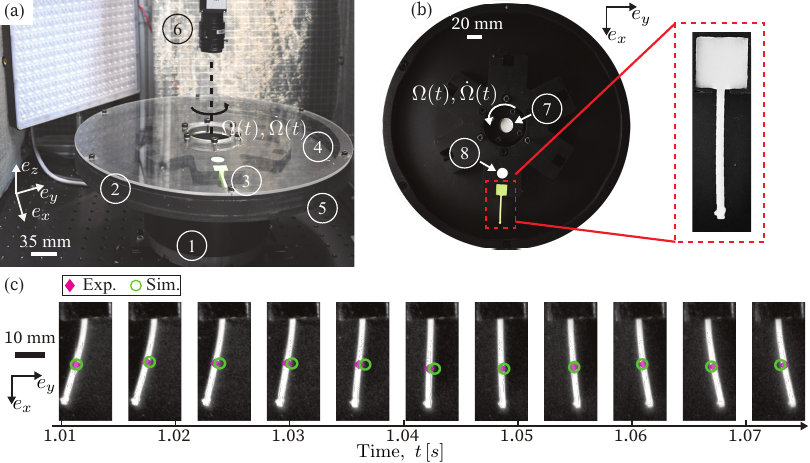}
    \caption{Experimental apparatus. 
    (a) Photograph of our setup: torque-controlled motor \textcircled{\footnotesize{1}}, rigid disk \textcircled{\footnotesize{2}}, cantilever beam \textcircled{\footnotesize{3}}, acrylic top plate \textcircled{\footnotesize{4}} and bottom dish \textcircled{\footnotesize{5}}, and camera \textcircled{\footnotesize{6}}. (b) Top view showing fixed \textcircled{\footnotesize{7}} and rotating \textcircled{\footnotesize{8}} markers for image stabilization. Inset: Zoom-in of the sample. (c) Rotating beam frames with overlaid symbols at midspan position: experiments (magenta diamonds) and simulations (green circles) for $\Omega_0=15\,\text{rad}/\text{s}$ and $f_d = 7\, \text{Hz}$.}
        \label{fig:Fig_2}
\end{figure*}

We validate the numerical model described in Sections~\ref{sec: Theoretical formulation} and \ref{sec: methodology - numerical} with experiments using the apparatus shown in \cref{fig:Fig_2}(a). The setup comprises a torque-controlled motor \textcircled{\footnotesize{1}} that rotates a rigid disk \textcircled{\footnotesize{2}} with the cantilevered beam \textcircled{\footnotesize{3}} mounted at a distance $R_o$ from the rotation center ($O$). A camera \textcircled{\footnotesize{6}} fixed in the lab frame images the beam deformation. The circular disk design balances the motor load, enabling higher velocities and accelerations while reducing high-frequency noise. Figure~\ref{fig:Fig_2}(b) shows a top view of the rotating disk with the clamped beam, while Fig.~\ref{fig:Fig_2}(c) presents representative experimental frames overlaid with the experimentally measured midspan of the beam (magenta diamonds) and simulation predictions (green circles) for driving parameters $\Omega_0=15\,\text{rad}/\text{s}$, and $f_d = 7\, \text{Hz}$. We now describe the apparatus components and protocol in detail.

\paragraph{Rotating system} Our rotating rigid disk (350\,mm diameter and 20\,mm thickness) is mounted to a torque-controlled motor (ETEL RTMBi140-150, driver AccurET 600). This motor accurately imposes arbitrary, time-varying velocity and acceleration profiles by using an input look-up table of angular position and time pairs, computed from \crefrange{eq: ang_vel}{eq: ang_acc}. An encoder records the angular position of the motor at 20\,kHz, providing high-precision velocity and acceleration measurements and real-time feedback for closed-loop control. The system's frameless architecture and high torque stability eliminate backlash and vibrations, enabling accurate tracking during dynamic operation.

\paragraph{Beam fabrication}
We cast beam samples from Elite Double 32, a vinyl polysiloxane (VPS) elastomer (Zhermack Dental) \cite{gutierrez-prieto_gyrophilia_2023}. This two-part system is mixed in a 1:1 mass ratio (base to catalyst) and poured into a mold to produce samples of width $b {=} 10\,\text{mm}$, thickness $h {=} 2\,\text{mm}$, and length $L {=} 40\,\text{mm}$. The material cures in 20 minutes, although its mechanical properties stabilize only after about 10 days. Previous studies \cite{baek2021finite,grandgeorge2021mechanics,johanns2021shapes} show that VPS32 behaves as an incompressible Neo-Hookean material, with a Young’s modulus of $E = 1.22 \pm 0.05\,\text{MPa}$ and a Poisson’s ratio of $\nu \approx 0.5$. The beam's first natural frequency is $f_n=6.37\,\text{Hz}$, computed via \cref{eq:natural_freq} with $\beta_1=1.875$. The beam parameters are summarized in \cref{table:params}; in subsequent sections, we refer to this as our \textit{reference beam}.

\begin{table}[ht]
  \centering
  
  \begin{tabular}{cccccc}
    \toprule
    $L\, [\text{mm}]$ & $h\, [\text{mm}]$ & $b\, [\text{mm}]$ & $E\, [\text{MPa}]$ &  $\rho\, [\text{kg}/\text{m}^3]$ \\
    \midrule
    40 & 2 & 10 & 1.22&  1170.2\\
    \bottomrule
  \end{tabular}
  
  \vspace{0.5ex}
  \caption{Geometric and material properties of the beam studied in \crefrange{sec: methodology - experiments}{sec:Defining the boundary between quasi-static and dynamic responses}.}\label{table:params}
\end{table}

\paragraph{Beam mounting} We clamp the cured beam inside a hollow acrylic disk, as shown in \cref{fig:Fig_2}(b). The disk comprises three parts: a central section that securely clamps the beam, and an acrylic top plate and bottom dish that enclose the beam. This enclosure reduces aerodynamic drag during rotation. The entire disk is painted black to enhance optical contrast for image processing. 

\paragraph{Imaging and image processing} We image the beam in the lab frame using a digital camera (\textit{IDS U3-3040SE-M-GL}) mounted 1.15\,m above the rotating disk. We process the acquired images using an in-house \textit{Python} algorithm based on the \texttt{skimage} library~\cite{van2014scikit}. This algorithm first stabilizes the images in the rotating frame by detecting the two circular markers visible in 
\cref{fig:Fig_2}(b): one fixed at the rotation center ($O$) and the other rotating at a known radius of 65\,mm. After detecting the markers, the algorithm rotates each frame using bi-cubic interpolation to align them. The algorithm then skeletonizes the beam to obtain its centerline. Since skeletonization may introduce artifacts near the beam's edges, we focus on the vertical deflection at the beam’s midspan ($y_m$), which is more robust to such errors. A representative series of frames is shown in \cref{fig:Fig_3}(b) with the detected midspan locations marked with magenta diamonds.

\paragraph{Experimental protocol} The beam is mounted onto the rigid rotating acrylic disk and enclosed between the two covers. A command is sent to the motor to rotate the disk at the desired angular velocity and acceleration profile (cf. Eqs.~\ref{eq: ang_vel}-\ref{eq: ang_acc}). A hardware signal from the motor controller triggers image acquisition, which continues throughout the experiment. Both the image and motor data are post-processed, and each experiment is repeated five times to quantify uncertainties. We explore the following range of drive parameters: frequency $f_d \in [1,8]\, \text{Hz}$ and mean angular velocity $\Omega_0 \in [0,60]\, \text{rad}/\text{s}$ (i.e., $\mathcal{C} \in [0.03,101]$ and $\mathcal{E} \in [0.2,1.4]$). These ranges are set by back-electromotive force limits during deceleration.


\section{Experimental validation of the numerical model}
\label{sec: Validation of Numerics Against Experiments}

We validate the numerical model against experiments for the reference beam (\cref{table:params}) by comparing time series and frequency response of the vertical midspan displacement, $y_m$, under different driving conditions.
\begin{figure}[h!]
\centering
\includegraphics[width=\columnwidth]{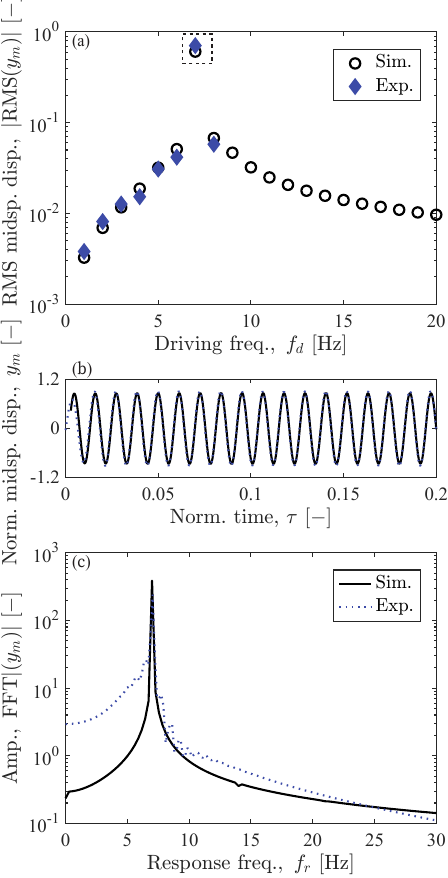}
\caption{Experimental validation of the numerical model (parameters in \cref{table:params}). (a) Resonance curve: RMS of dimensionless midspan displacement $\text{RMS}(y_m)$ versus drive frequency $f_d$ at mean angular velocity $\Omega_0 = 5\,\text{rad}/\text{s}$. Each experimental data point is the average of five repetitions; error bars are within the symbol size. (b) Time series of dimensionless midspan displacement, $y_m(\tau)$, at $f_d = 7\,\text{Hz}$. (c) Power spectrum of the time series in (b), showing resonance at 7\,Hz. Symbols/lines: (a) experiments (solid diamonds), simulations (open circles); (b,\,c) experiments (dashed), simulations (solid).
}
\label{fig:Fig_3}
\end{figure}

Figure~\ref{fig:Fig_3}(a) shows the root-mean-square value of the nondimensional midspan displacement,
\begin{equation}\label{eq:RMS}
 \text{RMS}(y_m) = \left(\frac{1}{N}\sum_{k=1}^{N} \left( y_{m,k} \right)^{2}\right)^{1/2},
\end{equation}
where $k$ is the index of the sampled point, versus drive frequency $f_d \in [1, 20]\,\text{Hz}$ at mean angular velocity $\Omega_0 = 5\,\text{rad}/\text{s}$. Experiments are limited to $f_d \leq 8\,\text{Hz}$ due to motor constraints (see \cref{sec: methodology - experiments}). The experimental and simulated $\text{RMS}(y_m)$ values show good agreement, both exhibiting a clear resonant peak at $f_d \approx 7\,\text{Hz}$, consistent with the first natural frequency of $f_n = 6.37\,\text{Hz}$ (see \cref{sec: methodology - experiments}). 

Figure~\ref{fig:Fig_3}(b) shows the time series of the dimensionless midspan displacement, $y_m(\tau)$, at the resonance drive frequency $f_0 = 7\,\text{Hz}$. Experiments and simulations match closely in both amplitude and frequency. The experimental curve represents the mean of five repetitions, with standard deviations smaller than the line width, indicating high reproducibility.

We also compute the power spectrum $|\text{FFT}(y_m)|$ of the midspan displacement time series. Figure~\ref{fig:Fig_3}(c) shows a representative power spectrum versus response frequency $f_r$ for the resonant case. Both experimental and numerical spectra confirm a resonance peak at $f_r \approx 7\,\text{Hz}$. The experimental spectrum exhibits higher broadband low-frequency noise, visible as an elevated baseline, due to parasitic vibrations in the apparatus.

The close agreement between experiments and simulations validates the numerical model for the parameter studies that follow.

\FloatBarrier

\section{Boundary between quasi-static and dynamic response}
\label{sec:Defining the boundary between quasi-static and dynamic responses}

Having validated the numerical model, we proceed to systematically explore the parameter space. We focus on the reference beam (parameters in \cref{table:params}) to characterize the boundary between quasi-static (non-inertial) and dynamic (inertial) response. To do so, we compare time series of the dimensionless midspan displacement, $y_m(\tau)$, from the full dynamical system \crefrange{eq:EOM1}{eq:EOM3} and the quasi-static approximation that neglects inertial effects. (In \cref{sec: Scaling law of the phase boundaries}, we will generalize this analysis for a broader range of the system's parameters.)

We introduce a criterion to distinguish between quasi-static and dynamic regimes based on the relative root-mean square (RMS) difference of $y_m(\tau)$. Using \cref{eq:RMS}, we define
\begin{equation}
    \begin{aligned}
     \text{RMS}_{\text{Dyn.}} {=} \text{RMS}(y^{\text{Dyn.}}_m) ~\text{ and }~ 
 \text{RMS}_{\text{QS}}   {=} \text{RMS}(y^{\text{QS}}_m)\,,
    \end{aligned}
\end{equation}
where $y_m^{\text{Dyn.}}$ and $y_m^{\text{QS}}$ are the dimensionless midspan displacements from the dynamical and quasi-static models. The relative RMS difference is
\begin{equation}\label{eq:Delta_RMS}
    \Delta{\text{RMS}} = \frac{|\text{RMS}_{\text{Dyn.}} - \text{RMS}_{\text{QS}}|}{\text{RMS}_{\text{Dyn.}}}.
\end{equation}
We classify the response as dynamic when $\Delta{\text{RMS}}\ge\Delta{\text{RMS}_{\text{th}}}$ and quasi-static otherwise, where $\Delta{\text{RMS}_{\text{th}}}$ is a threshold value. We set $\Delta{\text{RMS}_{\text{th}}} = 0.10$ but examine the boundary's sensitivity to this threshold in \cref{sec: Scaling law of the phase boundaries} and Appendix \ref{app:threshold}. 

\begin{figure*}[t!]
\centering
\includegraphics[width=0.85\textwidth]{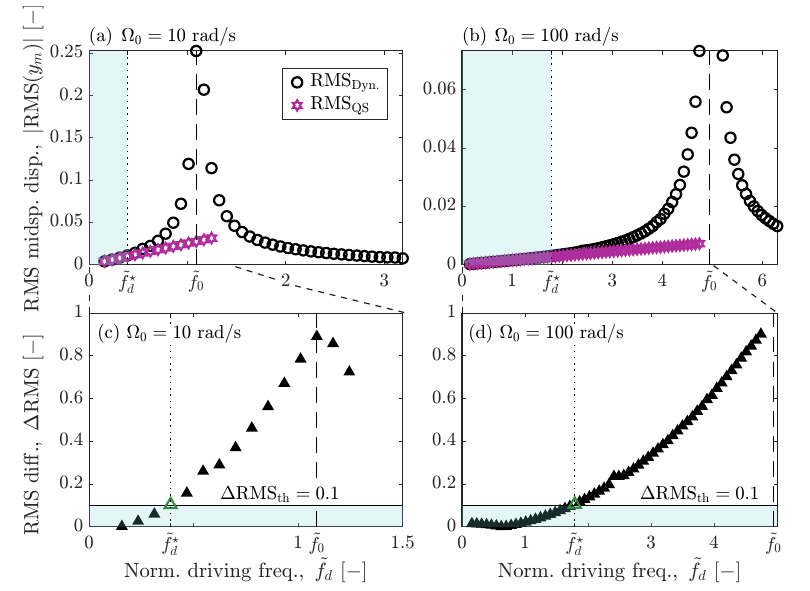}
\caption{Comparison of dynamical and quasi-static simulations for the reference beam. (a,\,b) $\text{RMS}(y_m)$ versus normalized drive frequency for $\Omega_0 = 10\,\text{rad/s}$ and $100\,\text{rad/s}$, respectively.  Circles: dynamical ($\text{RMS}_{\text{Dyn.}}$); stars: quasi-static ($\text{RMS}_{\text{QS}}$). (c,\,d) Relative RMS difference, $\Delta \text{RMS}$, for the same conditions. Vertical lines: critical normalized drive frequency $\tilde{f}_d^\star$ (dotted); normalized resonance frequency $\tilde{f}_0$ (dashed). Horizontal line in (c,\,d): threshold $\Delta \text{RMS}_\text{th}=0.1$. Open triangle: first point where $\Delta \text{RMS}\geq \Delta \text{RMS}_\text{th}$. Shaded region: quasi-static response; unshaded: dynamic response.}
\label{fig:Fig_4}
\end{figure*}

In \cref{fig:Fig_4}(a,b), we plot $\text{RMS}(y_m)$ versus the normalized drive frequency, $\tilde f_d=f_d/f_n$, where $f_n$ is the natural frequency defined in \cref{eq:natural_freq}, for the reference beam. Panels (a) and (b) show results for both the dynamical (circles) and quasi-static (stars) models at angular velocities $\Omega_0=10\,\text{rad/s}$ and $\Omega_0=100\,\text{rad/s}$, respectively. The dynamical model exhibits typical resonance (peak at $\tilde{f}_0=f_0/f_n$, represented by the vertical dashed line), whereas the quasi-static model yields nearly linear RMS versus $\tilde f_d$. The dynamical and quasi-static RMS are nearly indistinguishable at low drive frequencies but separate before resonance (in the dynamic case). The resonance frequency $\tilde{f}_0$ is higher for $\Omega_0=100\,\text{rad/s}$ than for $10\,\text{rad/s}$. Additionally, the agreement between the quasi-static and dynamical simulations extends to higher frequencies with increasing $\Omega_0$, which we explore next by examining $\Delta\text{RMS}$ in \cref{fig:Fig_4}(c,d). The open triangle, at the critical drive frequency $\tilde{f}_d^\star$ (vertical dotted line), marks the first point where $\Delta\text{RMS}\geq\Delta\text{RMS}_{\text{th}}$. For $\tilde{f}_d<\tilde{f}_d^\star$ (shaded region), the response is quasi-static, whereas for $\tilde{f}_d>\tilde{f}_d^\star$, the response is dynamic and the quasi-static approximation breaks down. 

We now systematically explore the drive parameter space to map the transition from quasi-static to dynamic behavior. 
Using \cref{eq:centrifugal_number,eq:euler_number}, we quantify the harmonic driving using the \textit{maximal} values of the centrifugal and Euler numbers:
\begin{align}
   &  \mathcal{C}_{\text{max}} =  (\Omega_0+a)^2\frac{\rho A R L^3}{EI} \label{eq:max C num} \quad \text{and} \\ 
   &  \mathcal{E}_{\text{max}} =  a 2\pi f_d\frac{\rho A  R L^3}{EI}. \label{eq:max E num}
\end{align}
Since the amplitude is fixed at $a=1\,\text{rad/s}$, varying the mean angular velocity, $\Omega_{0}$, modifies only $\mathcal{C}_{\max}$, whereas varying the drive frequency affects only $\mathcal{E}_{\max}$.

\begin{figure*}[t!]
        \centering
        \includegraphics[width=0.8\textwidth]{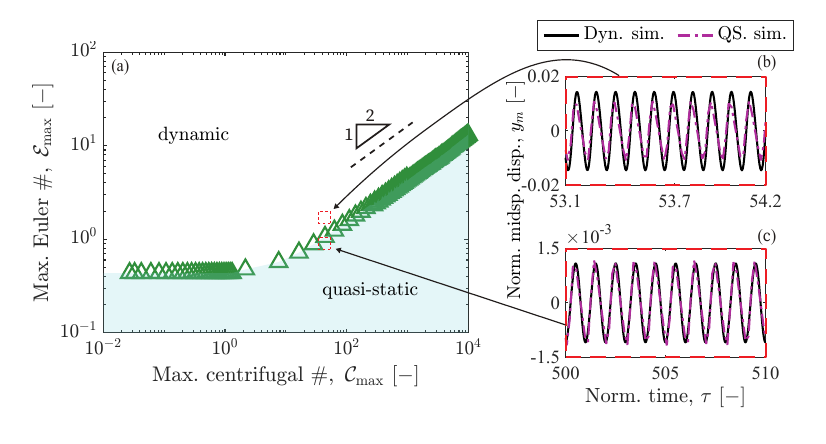}
        \caption{Phase diagram of quasi-static and dynamic regions for the reference beam. (a) Critical maximal Euler number $\mathcal{E}^\star$ (triangles) versus maximal centrifugal number $\mathcal{C}_{\text{max}}$ separates quasi-static (shaded) and dynamic (unshaded) regions. (b,\,c) Time series of dimensionless midspan displacement: dynamical (solid) and quasi-static (dashed) for $\mathcal{C}_{\text{max}} \approx 50$ with  (b) $\mathcal{E}_{\text{max}} \approx 2$ (above $\mathcal{E}^\star$) and (c) $\mathcal{E}_{\text{max}} \approx 1$ (below $\mathcal{E}^\star$).}
        \label{fig:Fig_5}     
    \end{figure*}

With these maximal values defined, we apply the RMS-based criterion to map the quasi-static-to-dynamic boundary in the $(\mathcal{C}_{\text{max}},\, \mathcal{E}_{\text{max}})$ plane, as plotted in \cref{fig:Fig_5}(a). Each triangle denotes 
the first pair of simulations (dynamical and quasi-static) for which $\Delta \text{RMS}\geq 0.10$. The data are obtained by fixing $\mathcal{C}_{\text{max}}$ (equivalently, $\Omega_0$) and running both dynamical and quasi-static simulations while increasing $\mathcal{E}_{\text{max}}$ by linearly increasing $f_d$. From the critical drive frequency, $\tilde{f}_d^\star$ (see \cref{fig:Fig_4}), we obtain the critical maximal Euler number, $\mathcal{E}^\star$, through 
\cref{eq:max E num}. The $\mathcal{E}^\star (\mathcal{C}_{\text{max}})$ curve partitions the parameter space into two regions: a quasi-static region (lower, shaded), where inertial effects are negligible, and a dynamic region (upper, unshaded), where inertial effects dominate. For $\mathcal{C}_{\text{max}}\lesssim 10$, the critical maximal Euler number remains nearly constant at $\mathcal{E}^\star\approx5$, after which it scales as $\mathcal{E}^\star \sim \mathcal{C}_{\text{max}}^{1/2}$. 

The shape of this $\mathcal{E}^\star (\mathcal{C}_{\text{max}})$ boundary reveals two key behaviors. For a given Euler number, increasing the centrifugal number (moving right in the phase diagram) eventually produces quasi-static response for sufficiently large $\mathcal{C}_{\text{max}}$. Conversely, for a given centrifugal number, increasing the Euler number (moving upwards in the phase diagram) eventually produces dynamic response with non-negligible inertial effects for sufficiently large $\mathcal{E}_{\text{max}}$.

To illustrate the breakdown of the quasi-static approximation in the dynamic regime, \cref{fig:Fig_5}(b,\,c) 
plots representative time series of the dimensionless midspan displacement for two parameter pairs: $(\mathcal{C}_{\text{max}},\, \mathcal{E}_{\text{max}})=(50,\,2)$ and $(50,\,1)$, corresponding to locations just above and below the $\mathcal{E}^\star$ boundary, respectively. As expected from their location relative to the boundary, the dynamical and quasi-static simulations disagree in panel (b) (dynamic regime), whereas they agree well in panel (c) (quasi-static regime).

\begin{figure}[h!]
    \centering
    \includegraphics[width=\columnwidth]{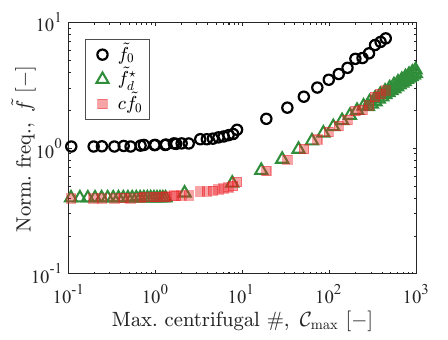}
    \caption{Normalized resonance frequency $\tilde{f}_0$ (circles), critical normalized drive frequency $\tilde{f}_d^\star$ (triangles), and the relation $\tilde{f}_d^\star=c\tilde{f}_0$ with constant $c < 1$ (squares) versus centrifugal number $\mathcal{C}_{\text{max}}$ from dynamical simulations of the reference beam. The constant is determined to be $c=0.385 \pm 0.001$ through a nonlinear least-squares fit, as described in Section~\ref{sec: Scaling law of the phase boundaries}.} 
    \label{fig:Fig_6}     
\end{figure}

The square-root scaling $\mathcal{E}^\star \sim \mathcal{C}_{\text{max}}^{1/2}$ observed in the phase boundary is related to how centrifugal loading affects the beam's resonance frequency. In Fig.~\ref{fig:Fig_6},
we plot the normalized resonance frequency $\tilde{f}_0$, the critical frequency $\tilde{f}_d^\star$, and the linear relation $\tilde{f}_d^\star=c\tilde{f}_0$ as functions of $\mathcal{C}_{\text{max}}$. The proportionality constant is determined to be $c=0.385 \pm 0.001$ by nonlinear least-squares fitting using the MATLAB function \texttt{lsqcurvefit} (see \cref{sec: Scaling law of the phase boundaries}). The results in \cref{fig:Fig_6} suggest that the critical frequency corresponds to the resonance frequency scaled by the factor $c$. 

For $\mathcal{C}_{\text{max}} \lesssim  8$, the resonance frequency remains constant at $\tilde{f}_0 \approx 1$, whereas for $\mathcal{C}_{\text{max}} \gtrsim 8$, it scales as $\tilde{f}_0 \sim \mathcal{C}_{\text{max}}^{1/2}$. The critical frequency $\tilde{f}_d^\star$ exhibits qualitatively similar behavior. This increase in frequency with angular velocity is attributed to the classical centrifugal stiffening effect: rotation induces axial tension in the beam, thereby increasing its natural frequency.  \cite{wright1982vibration} derived exact solutions for centrifugally stiffened beams; \cite{yoo_vibration_1998} reported natural frequency changes versus angular velocity and geometry; \cite{behzad2004effect} demonstrated how rotation-induced axial force affects shaft frequency. These studies validate the classical Southwell relation \cite{southwell_free_1921} and its associated square-root scaling, which our rotating system also exhibits. This scaling proves central to the phase boundary in \cref{fig:Fig_5} and motivates the theoretical analysis in the next section. 

\section{Generalized theoretical description of the phase boundary}
\label{sec: Scaling law of the phase boundaries}

Having established the transition between quasi-static and dynamic behavior for our reference beam (\cref{table:params}), we now develop a theoretical description for the phase boundary between these regimes. The starting point for our analysis is the premise that the physical mechanism driving this boundary is the well-known centrifugal stiffening effect \cite{southwell_free_1921}. For ease of notation, we define $G=\rho A R L^3/EI$, combining geometric and mechanical parameters, such that
\begin{align}
    \mathcal{E}_\text{max} &= Ga2\pi f_d, \quad \text{and} \label{eq: Euler max} \\
    \mathcal{C}_\text{max} &= G(\Omega_0+a)^2. \label{eq: cen max}
\end{align}
We recall that the critical maximal Euler number is
\begin{equation}\label{eq:critical_max_Euler}
    \mathcal{E}^\star = G a 2\pi f^\star_d.
\end{equation}
Furthermore, aligning with the data in \cref{fig:Fig_4}, we posit that the critical driving frequency follows
\begin{equation}\label{eq:cf0}
    f_d^\star = c f_0,
\end{equation}
where $f_0$ is the resonance frequency and $c<1$ is a proportionality constant. While the critical ratio $c=f^\star_d/f_0$ is commonly used in the literature as a heuristic criterion for the quasi-static/dynamic boundary \cite{piersol2010harris,chopra2012dynamics}, in Appendix \ref{app:second_order_system_response}, using a canonical second-order dynamical system, we show that $c$ is set by the chosen error threshold and remains constant for sufficiently underdamped systems. Given the underdamped nature of our system, the theoretical argument presented in Appendix \ref{app:second_order_system_response} rationalizes the constant proportionality posited in \cref{eq:cf0} and observed in the data plotted in \cref{fig:Fig_4}.

The classical Southwell equation determines that rotation shifts the resonance frequency due to centrifugal stiffening. Following \cite{southwell_free_1921}, the maximal shifted resonance frequency of the first vibrational mode for a beam rotating with angular velocity $\Omega$ is
\begin{equation}\label{eq: centrifugal_stiffening}
    (2\pi f_0)^2 = (2\pi f_n)^2 + S_1\Omega^2.
\end{equation}
The first Southwell coefficient $S_1$ (a geometric parameter dependent on the beam's fundamental mode shape) \cite{bazoune2005relationship} is computed using an established analytical expression derived from the Rayleigh quotient, defined as the ratio of the potential energy due to centrifugal tension to the kinetic energy of the first mode \cite{southwell_free_1921,lo1952bending,meirovitch1967analytical}.
Substituting the maximal angular velocity in our system ($\Omega_0+a$) into \cref{eq: centrifugal_stiffening} we find the maximal resonance frequency
\begin{equation}\label{eq:Southwell}
    (2\pi f_0)^2 = (2\pi f_n)^2 + S_1(\Omega_0+a)^2.
\end{equation}
Substituting \cref{eq:Southwell,eq:cf0} into \cref{eq:critical_max_Euler} yields
\begin{equation}\label{eq:sub1}
    \begin{aligned}
    (\mathcal{E}^\star)^2 &= (Ga2\pi c f_0)^2 \\ &= (Gac)^2\left[(2\pi f_n)^2 + S_1(\Omega_0+a)^2\right],
    \end{aligned}
\end{equation}
which we then rewrite in terms of the centrifugal and Euler numbers:
\begin{equation}
    \begin{aligned}
    (\mathcal{E}^\star)^2 
    = c^2\left(\mathcal{E}_n^2+S_1Ga^2\mathcal{C}_\text{max}\right),
     \end{aligned}
\end{equation}
where $\mathcal{E}_n$ denotes the maximal Euler number at $f_d=f_n$. Since we are interested only in the positive solutions, we write
\begin{equation}\label{eq:analytical_sol}
    \mathcal{E}^\star = c\sqrt{\mathcal{E}_n^2+S_1Ga^2\mathcal{C}_\text{max}}.
\end{equation}

We validate Eq.~(\ref{eq:analytical_sol}) by conducting numerical simulations over a broad range of material and geometric parameters in the following ranges: Young's modulus $E \in [10^5,\,10^9]\,\text{Pa}$, beam length $L \in [10^{-4},\,10^{-2}]\,\text{m}$, and thickness-to-length slenderness ratio $\lambda_1=h/L \in [10^{-1},\,10^{-2}]$.

\begin{figure}[h!]
    \centering
    \includegraphics[width=\columnwidth]{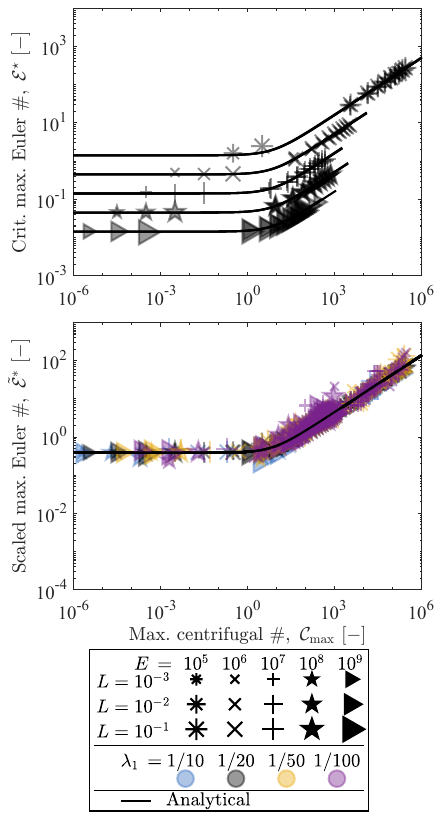}
    \caption{The quasi-static/dynamic boundary for varying Young’s moduli, lengths, and slenderness ratios. (a) The critical maximal Euler number $\mathcal{E}^\star$, and (b) the scaled maximal Euler number $\tilde{\mathcal{E}}^\star$ as functions of the maximal centrifugal number $\mathcal{C}_{\max}$. Symbols denote different values of Young’s modulus; larger marker sizes indicate longer beams; and colors indicate slenderness ratio, with $\lambda_1 = 1/20$ in (a). The analytical solutions from \cref{eq:analytical_sol} and \cref{eq:scaled_analytical_sol} are shown as black solid lines in (a) and (b), respectively.}
    \label{fig_7}
\end{figure}

In \cref{fig_7}(a), we plot the critical maximal Euler number versus maximal centrifugal number $\mathcal{E}^\star(\mathcal{C}_\text{max})$ for different values of $E$ and $L$, and with $\lambda_1 = 1/20$, maintaining the threshold $\Delta\text{RMS}_{\text{th}}=0.10$. The predictions from \cref{eq:analytical_sol} (black solid line) are in excellent agreement with the numerical data. Different values of $E$ introduce a multiplicative pre-factor that shifts the data vertically in the log-log plot. Since this shift stems from the beams' differing natural frequencies, we rescale the numerical results and \cref{eq:analytical_sol} by $1/\mathcal{E}_n$, yielding the scaled maximal Euler number:
\begin{equation}\label{eq:scaled_analytical_sol}
    \tilde{\mathcal{E}}^\star = \frac{\mathcal{E}^\star}{\mathcal{E}_n} = c\sqrt{1+\frac{S_1Ga^2}{\mathcal{E}_n^2}\mathcal{C}_\text{max}},  
\end{equation}
where, we recall that $c=0.385 \pm 0.001$. In \cref{fig_7}(b), we plot the scaled quantity for the full range of $E$, $L$, and $\lambda_1$ explored. The analytical expression in \cref{eq:scaled_analytical_sol} (black solid line) collapses all data onto a master curve, confirming that the phase boundary is fully described by the natural frequency shift due to centrifugal stiffening.

\FloatBarrier
\section{Conclusions and future work}
\label{sec: conclusions}

We adapted an \textit{elastica}-based reduced-order model and validated it against precision experiments to simulate rotating cantilever beams driven harmonically. Systematically exploring the parameter space, we identified two distinct regimes: one quasi-static, where inertial effects remain negligible under unsteady harmonic driving, and another dynamic, where inertia significantly influences the response. We characterized the boundary between these regimes using the Euler and centrifugal numbers, which quantify centrifugal and Euler forces relative to characteristic bending forces. The critical Euler number remains approximately constant at low centrifugal numbers but transitions to square-root scaling with centrifugal number at higher rotation rates. 

We derived a predictive theoretical description for this transition, revealing that centrifugal stiffening governs the scaling. Counterintuitively, faster rotation produces less dynamic response: increasing rotation raises the beam's resonance frequency, making the beam effectively stiffer and expanding the quasi-static regime to higher drive frequencies.

Our results establish the conditions under which a quasi-static approximation remains valid for rotating flexible beams under unsteady loading. This framework can be applied to the design and control of robotic arms, turbine blades, soft actuators, and deployable aerospace components. 

Future work should investigate more complex nonlinear phenomena such as modal interactions and potential instabilities. Preliminary observations suggest the system exhibits complex dynamics, potentially including chaos, under certain driving conditions, warranting further investigation.

\appendix

\section{Discretized equations of motion} \label{app:A}

We define $N+1$ uniform grid points along the dimensionless arclength $\ads_i = i \Delta s$ with $\Delta \ads = 1/N$. At each grid point, we define discretized coordinates $\theta_i =  \theta(\ads_i)$, $\adx_i =  \adx(\ads_i)$, $\ady_i =  \ady(\ads_i)$, and solve for $\theta_i$ to avoid enforcing the inextensibility constraint at each step. Using the discrete version of the kinematic relations in \cref{eq:x_s_nad_x_y}, the discrete Cartesian coordinates $(\adx_i,\,\ady_i)$ are computed with the trapezoid rule. For every $\adx_i $ with $ i=1, ... ,N{-}1$ we get
\begin{equation}
    \adx_i = \frac{\Delta\ads}{2}\sum_{k=1}^i{\cos\theta_{k-1} + \cos\theta_{k}}\,
\end{equation}
and conducting this integration for every $i$ yields the linear system
\begin{equation}\label{eq: numX}
\begin{split}
    \mathbf{\adx} &= \begin{bmatrix} \adx_1 & \adx_2 & \cdots & \adx_{N{-}1}\end{bmatrix}^\text{T} \\& = \left(\frac{\Delta \ads}{2}+\mathbf{L}\cdot\mathbf{\tilde{c}}\right),
\end{split}
\end{equation} 
where $\mathbf{L}$ is the lower triangular matrix 
\begin{equation}\label{eq: L}
    \mathbf{L} = \frac{\Delta\ads}{2}
            \begin{pmatrix}
            1      & 0       & 0      & \cdots  & 0\\
            2      & 1       & 0     & \cdots  & 0\\
            2      & 2       & 1    & \cdots  & 0\\
            \vdots & \vdots &  \vdots &  \vdots      & \vdots\\
            2      & 2       & 2    & \cdots       & 1 \\
            \end{pmatrix},
\end{equation}
and 
\begin{equation}
    \mathbf{\tilde{c}} = 
    \begin{bmatrix}
        \cos{\theta_1} & \cdots & \cos{\theta_{N{-}1}}
    \end{bmatrix}^\text{T}.
\end{equation}
Similarly, for the $y$-coordinate, we find 
\begin{equation}\label{eq: numY}
    \mathbf{\ady} = \begin{bmatrix} \ady_1 & \ady_2 & \cdots & \ady_{N{-}1}\end{bmatrix}^\text{T} =  \left(\mathbf{L}\cdot\mathbf{\tilde{s}}\right),
\end{equation} 
where 
\begin{equation}
    \mathbf{\tilde{s}} = 
    \begin{bmatrix}
        \sin{\theta_1} & \cdots & \sin{\theta_{N{-}1}}
    \end{bmatrix}^\text{T}.
\end{equation}
The first and second time derivatives of $\mathbf{\adx}$ and $\mathbf{\ady}$ are
\begin{equation} \label{eq: Dervs}
    \begin{split}    
        &\dot{\mathbf{\adx}} = -\mathbf{L}\left(\tilde{\mathbf{S}}_1 \cdot\dot{\mathbf{\Theta}}\right),
        \\
        &\ddot{\mathbf{\adx}} = -\mathbf{L}\left(\tilde{\mathbf{S}}_1 \cdot\ddot{\mathbf{\Theta}} + \tilde{\mathbf{C}}_1 \cdot\dot{\mathbf{\Theta}}^2\right),
        \\
        &\dot{\mathbf{\ady}} = \mathbf{L}\left(\tilde{\mathbf{C}}_1 \cdot\dot{\mathbf{\Theta}}\right),
        \\
        &\ddot{\mathbf{\ady}} = \mathbf{L}\left(\tilde{\mathbf{C}}_1 \cdot\ddot{\mathbf{\Theta}} - \tilde{\mathbf{S}}_1 \cdot\dot{\mathbf{\Theta}}^2\right),
    \end{split}  
\end{equation}
where $\dot{\mathbf{\Theta}}$ and $\ddot{\mathbf{\Theta}}$ are the first and second time derivatives of $\mathbf{\Theta} = \left[ \theta_1,...,\theta_{N{-}1} \right ]$, and $\dot{\mathbf{\Theta}}^2 = \left[ \dot{\theta}_{1}^2,...,\dot{\theta}_{N{-}1}^2 \right ]$.
In \cref{eq: Dervs},  $\tilde{\mathbf{S}}_1$ and $\tilde{\mathbf{C}}_1$ are the diagonal matrices: 
\begin{align}
\setlength{\arraycolsep}{2pt} 
\tilde{\mathbf{S}}_1 &=
\begin{pmatrix} 
    \sin\theta_1 & 0 & \cdots & 0 \\
    0 & \sin\theta_2 & \cdots & 0 \\
    \vdots & \vdots & \ddots & 0 \\
    0 & 0 & \cdots & \sin\theta_{N{-}1}
\end{pmatrix}
,
\label{eq:S1}
\\[1ex]
\tilde{\mathbf{C}}_1 &=
\begin{pmatrix} 
    \cos\theta_1 & 0 & \cdots & 0 \\
    0 & \cos\theta_2 & \cdots & 0 \\
    \vdots & \vdots & \ddots & 0 \\
    0 & 0 & \cdots & \cos\theta_{N{-}1}
\end{pmatrix}
.
\label{eq:C1}
\end{align}
Enforcing Neumann boundary condition $ \theta(1,\adtau)_{,s} = 0$ with the backwards Euler method we obtain 
\begin{equation}\label{eq: Neumann BC}
    \frac{\theta_{N}-\theta_{N-1}}{\Delta\ads} = 0.
\end{equation}
Applying \cref{eq: Neumann BC} together with the boundary condition $\theta_0 = 0$ at the clamped end, we simplify the system to solve for $N{-}1$ grid points.

We now discretize the tension equations in \crefrange{eq:EOM1}{eq:EOM2} with zero tension at the free end, $\ads = 1$.
\begin{equation}\label{eq: NumT}
    \begin{split}
        n_x &= - \int_{\ads}^{1} \bigg(
            \ddot{x} + \adeta \dot{x} - 2\mathcal{I}\dot{y} -{}\\
        &\mkern50mu\left[\mathcal{C}\left(1 - \addelta + \addelta\adx \right)
            + \addelta\mathcal{E}\ady\right]\bigg)\,\text{d}s
        \\
        n_y &= - \int_{\ads}^{1} \bigg(
            \ddot{y} + \adeta \dot{y} + 2\mathcal{I}\dot{y} +{}\\
        &\mkern50mu\left[\mathcal{E}\left(1 - \addelta + \addelta\adx \right)
            - \addelta\mathcal{C}\ady\right]\bigg)\,\text{d}s
    \end{split}
\end{equation}
Substituting \cref{eq: numX,eq: numY,eq: Dervs} into \cref{eq: NumT} and rearranging yields
\begin{align} 
    \begin{split}\label{eq: Disc_tension_x}      
        n_x &= \mathcal{C}\left[1 + \left(\addelta - 1\right)\mathbf{\ads}-\addelta\frac{\mathbf{\ads}^2}{4}\right] +\\&
        \mkern50mu \mathbf{U}\mathbf{L}\Bigl[\tilde{\mathbf{S}}_1 \ddot{\mathbf{\Theta}} + \tilde{\mathbf{C}}_1 \dot{\mathbf{\Theta}}^2 +\\&
        \mkern90mu \left(\adeta \tilde{\mathbf{S}}_1+2\mathcal{I}\tilde{\mathbf{C}}_1\right)\dot{\mathbf{\Theta}} +\\&
        \mkern150mu \addelta\left(\mathcal{C}\mathbf{\tilde{c}} + \mathcal{E}\mathbf{\tilde{s}}\right)\Bigl]
   \end{split}
        \\
    \begin{split} \label{eq: Disc_tension_y}
        n_y &=  -\mathcal{E}\left[1 + \left(\addelta - 1\right)\mathbf{\ads}-\addelta\frac{\mathbf{\ads}^2}{4}\right] - \\&
        \mkern50mu \mathbf{U}\mathbf{L}\Bigl[\tilde{\mathbf{C}}_1 \ddot{\mathbf{\Theta}} - \tilde{\mathbf{S}}_1 \dot{\mathbf{\Theta}}^2 +\\&
        \mkern90mu \left(\adeta \tilde{\mathbf{C}}_1 - 2\mathcal{I}\tilde{\mathbf{S}}_1\right)\dot{\mathbf{\Theta}}  +\\&
        \mkern150mu \addelta\left(\mathcal{E}\mathbf{\tilde{c}} - \mathcal{C}\mathbf{\tilde{s}}\right) \Bigr],
    \end{split}
\end{align}
where $\mathbf{U}$ is the upper triangular matrix
\begin{equation}\label{eq: U}
    \mathbf{U} = \frac{\Delta\ads}{2}
    \begin{pmatrix}
    1      & 2   & 2     & \cdots  & 3\\
    0      & 1      & 2      & \cdots  &  3\\
    0      & 0      & 1      & \cdots  &  3\\
    \vdots & \vdots  &  \vdots & \vdots & \vdots\\
    0      & 0      & 0      & \cdots       & 2\\
    \end{pmatrix}
\end{equation}
and $\mathbf{\ads} = \left[\ads_1,...,\ads_{N{-}1}\right]$ is the discretized arclength.
Next, we discretize the moment \cref{eq:EOM3}. Using center differences, applying the boundary conditions and \cref{eq: Neumann BC} we find 
\begin{equation}\label{eq: Discrete_theta_ss}
    {\theta}_{,s} = -\mathbf{K}\mathbf{\Theta},     
\end{equation}
where $\mathbf{K}$ is the stiffness matrix 
\begin{equation} \label{eq: K}
    \mathbf{K} = -\frac{1}{\Delta \ads}
    \begin{pmatrix}
        -2      &  1      &  0  &  0   & \cdots &  0 \\
         1      & -2      &  1  &  0   & \cdots &  0 \\
         0      &  1      & -2  &  1   & \cdots &  0 \\
         \vdots &  \vdots & \vdots & \vdots & \vdots   &  \vdots \\
         \vdots &  \vdots & \vdots & 1 & -2   &  1 \\
         0      &  0      & 0   & 0    & 1      & -1 \\
    \end{pmatrix}.
\end{equation}

Substituting \crefrange{eq: Disc_tension_x}{eq: Discrete_theta_ss} into \crefrange{eq:EOM1}{eq:EOM3}, applying trigonometric identities, and rearranging, we obtain the discretized equations of motion in \cref{eq:discretizedEq} in terms of the tangent angle $\theta_i$, which we rewrite as:
\begin{equation}
        \mathbf{M} \ddot{\mathbf{\Theta}} + \mathbf{C} \dot{\mathbf{\Theta}} + \mathbf{K}\mathbf{\Theta} = \mathbf{f} .
\end{equation}
The mass matrix $\mathbf{M}$ decomposes as $\mathbf{M} = \mathbf{U}\mathbf{L}\mathbf{\tilde{C}}$, where $\mathbf{U}$ and $\mathbf{L}$ are the upper and lower triangular matrices in \cref{eq: L,eq: U}.
The matrix $\mathbf{\tilde{C}}$ encodes cosines of tangent-angle differences $\Delta\theta_{i,j} = \theta_j - \theta_i$ between nodes $i$ and $j$. To simplify notation we define $c_{i,j} = \cos{\Delta\theta_{i,j}}$, and $s_{i,j} = \sin{\Delta\theta_{i,j}}$:
\begin{equation}
\mathbf{\tilde{C}} =
\begin{pmatrix}
    1 & c_{1,2} & c_{1,3} & \cdots & c_{1,N{-}1} \\
    c_{1,2} & 1 & c_{2,3} & \cdots & c_{2,N{-}1} \\
    c_{1,3} & c_{2,3} & 1 & \cdots & c_{3,N{-}1} \\
    \vdots & \vdots & \vdots & \ddots & \vdots \\
    c_{1,N{-}1} & c_{2,N{-}1} & c_{3,N{-}1} & \cdots & 1
\end{pmatrix}.
\end{equation}
The damping matrix is $\mathbf{C} = \adeta \mathbf{M}+2\mathcal{I}\mathbf{B}$. Here, $\mathbf{B} = \mathbf{U}\mathbf{L}\mathbf{\tilde{S}}$ is the gyroscopic matrix, which represents the Coriolis forces acting on the beam through the nondimensional inertial number $\mathcal{I}$. This matrix arises from the spatial integration of geometric coupling terms defined by
\begin{equation}
    \mathbf{\tilde{S}} = 
    \begin{pmatrix}
        0       & s_{1,2} & s_{1,3} & \cdots & s_{1,N{-}1} \\
        s_{1,2} & 0       & s_{2,3} & \cdots & s_{2,N{-}1} \\
        s_{1,3} & s_{2,3} & 0       & \cdots & s_{3,N{-}1} \\
        \vdots  & \vdots  & \vdots  & \ddots & \vdots      \\
        s_{1,N{-}1} & s_{2,N{-}1} & s_{3,N{-}1} & \cdots & 0
    \end{pmatrix}.
\end{equation}
However, while the Coriolis term is formally included in the derivation,its effects are ultimately shown to be negligible.
The forcing term $\mathbf{f}$ contains the gyroscopic nonlinearities $\dot{\mathbf{\Theta}}^2$ and trigonometric nonlinearities from the diagonal matrices $\tilde{\mathbf{S}}_1, \tilde{\mathbf{C}}_1$ in \cref{eq:S1,eq:C1},
and decomposes as
\begin{equation}
\begin{split}
    \mathbf{f} &= -\mathbf{B}\dot{\mathbf{\Theta}}^2 - \mathbf{U}\mathbf{L}\addelta\mathcal{E}\mathbf{e} \\&
        \mkern77mu+ \left[1+\left(\addelta-1\right)\mathbf{\ads} - \addelta\frac{\mathbf{\ads}^2}{4}\right]\\&
        \mkern98mu\left(\mathcal{E}\mathbf{\tilde{C}}_1 - \mathcal{C}\mathbf{\tilde{S}}_1\right)\,. 
\end{split}
\end{equation}
This term appears in the right hand side of \cref{eq:discretizedEq}.
 
\section{Damping coefficient} \label{app:B}

The damping coefficient of an underdamped unforced system is $\gamma =|2\rho A/t_{\text{decay}}|$ \cite{strogatz2018nonlinear}. We determine the damping coefficient by perturbing the beam from equilibrium and measuring a time series of the horizontal deflection at midspan. The decay time $t_{\text{decay}}$ is extracted by fitting the exponential function $A \exp{(-t/t_{\text{decay}})}$ to the signal. We find that $\gamma = 0.039 \pm 0.008$. For the dimensionless \crefrange{eq:EOM1}{eq:EOM3}, we nondimensionalize the damping coefficient as $\eta=\gamma L^2/\sqrt{\rho AEI}$.

\section{Convergence test for grid points $N$} \label{app:C}

\begin{figure}[h!]
        \centering
        \includegraphics[width=\columnwidth]{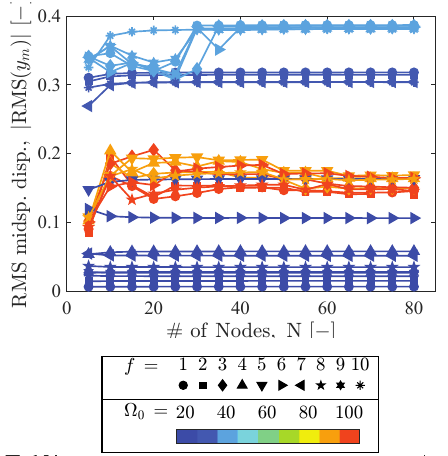}
        \caption{Convergence test for the number of grid points $N$ using drive frequencies $f_d \in [1,10]\,\text{Hz}$ and mean angular velocities $\Omega_0 \in [10,100]\,\text{rad/s}$.}
        \label{fig:B.8}
    \end{figure}

We determine the appropriate number of grid points, $N$, through convergence tests using drive frequencies $f_d \in [1,10]\,\text{Hz}$ and mean angular velocities $\Omega_0 \in [10,100]\,\text{rad/s} $. We compute the RMS of the dimensionless midspan displacement $\text{RMS}(y_m)$ (see \cref{eq:RMS}) for $N \in [5,80]$ grid points. Figure~\ref{fig:B.8} shows that for $N \gtrsim 40$, the RMS amplitude converges.
            
    \FloatBarrier  
        
\section{Effects of threshold $\Delta \text{RMS}_{\text{th}}$} \label{app:threshold}

\begin{figure}[h!]
        \centering
        \includegraphics[width=\columnwidth]{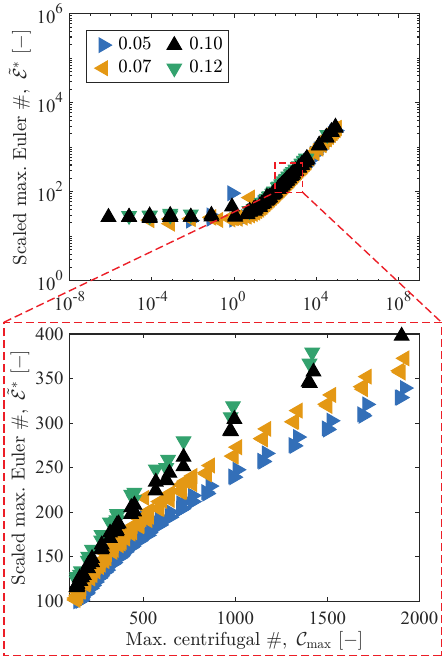}
        \caption{Quasi-static/dynamic boundary for varying Young’s moduli, lengths, and $\Delta \text{RMS}_{\text{th}}$. 
        (a) Scaled maximal Euler number $\tilde{\mathcal{E}}^\star$ versus the maximal centrifugal number $\mathcal{C}_{\max}$. 
        (b) Magnified version of panel (a), focusing on the range $\tilde{\mathcal{E}}^\star \in [100,400],\,\mathcal{C}_{\max} \in [100,2000]$. Blue right-pointing, orange left-pointing, black upward, and green downward triangles correspond to $\Delta \text{RMS}_{\text{th}} = 0.05,\,0.07,\,0.10,\,0.12$, respectively.}
        \label{fig:D.9}
    \end{figure}

We investigate the effect of the threshold $\Delta\text{RMS}_{\text{th}}$ by repeating the numerical simulations from \cref{sec: Scaling law of the phase boundaries} with $\Delta \text{RMS}_{\text{th}}= 0.05,\,0.07,\,0.10,\,0.12$, and $\lambda_1 = 1/20$. The resulting boundaries are shown in \cref{fig:D.9}. As $\Delta \text{RMS}_{\text{th}}$ increases, the curves shift slightly upward in the log-log plot, indicating that the multiplicative constant $c$ that determines the plateau value $\tilde{\mathcal{E}}^\star_0 = (\pi\lambda_1AL^3\sqrt{\rho}/\beta_1^2I\sqrt{3E})cf_n$ in the limit $\mathcal{C}_{\max}\rightarrow 0$ depends on the chosen $\Delta\text{RMS}_{\text{th}}$. The extracted constants are $c = 0.311\pm0.001,\,0.345\pm0.001,\,0.385\pm0.001,\,0.404\pm0.001$ for $\Delta \text{RMS}_{\text{th}} = {0.05,\,0.07,\,0.10,\,0.12}$, respectively. Throughout the main text, we use the threshold value $\Delta \text{RMS}_{\text{th}} = 0.10$ (black upward triangles). 
    
\FloatBarrier

\section{Quasi-static condition of the canonical second-order dynamical system} \label{app:second_order_system_response}

We apply classic dynamical systems theory to relate the threshold criterion ($\Delta\text{RMS}_\text{th}$) to the frequency ratio $c_0 = f/f_n$ defined in \cref{eq:cf0}. Consider the viscously damped, harmonically forced canonical second-order nondimensional dynamical system \cite{inman2009engineering}:
\begin{equation} \label{eq:2nd order system}
    \frac{\text{d}^2u}{\text{d}\tau^2} + 2\zeta\frac{\text{d}u}{\text{d}\tau} + u = \frac{F_0}{k}\cos\left(c_0\tau\right)
\end{equation}
where $\zeta$ is the nondimensional viscous damping ratio, $F_0$ is the drive amplitude, and $k$ represents the system stiffness.
\Cref{eq:2nd order system} governs the motion of various driven physical systems \cite{rao2004mechanical}. \Cref{table:systems} summarizes the physical interpretation of these generalized constants for a driven mass-spring system, a pendulum, and a cantilevered elastic beam. In this table, $k_s$ denotes the linear spring constant, $m$ the mass, $g$ the gravitational acceleration, and $L$ the characteristic length. Following the definitions in the main text, $E$, $I$, $\rho$, $A$, $\gamma$, and $\beta_1$ represent Young's modulus, area moment of inertia, density, cross-sectional area, the viscous damping coefficient, and the modal constant, respectively.

\begin{table}[h!]
  \centering
  
  \begin{tabular}{lccc}
    \toprule
    & $\zeta$ & $k$ & $f_n$ \\
    \midrule
    Mass-spring & $\gamma / (2\sqrt{k_sm})$ & $k_s$ & $\sqrt{k_s/m}$ \\
    Pendulum    & $\gamma / (2mL\sqrt{gL})$ & $mgL$ & $\sqrt{g/L}$ \\
    Cantilever beam        & $\gamma L^2 / (2\beta_1^2\sqrt{EI \rho A})$ & $3EI/L^3$ & $\beta_1^2 \sqrt{EI / (\rho A L^4)}$ \\
    \bottomrule
  \end{tabular}
  
  \vspace{0.5ex}
  \caption{Physical interpretation of nondimensional parameters for a driven mass-spring, pendulum, and cantilever beam. All are described by \cref{eq:2nd order system}.}\label{table:systems}
\end{table}

The steady-state amplitudes of the dynamic and quasi-static responses, $A_\text{dyn}$ and $A_\text{QS}$ respectively, are derived from the magnitude of the complex exponential solution to \cref{eq:2nd order system}, and are given by \cite{french2017vibrations}:
\begin{align}
        &A_\text{dyn} = \frac{F_0/k}{\sqrt{(1-c_0^2)^2+(2\zeta c_0)^2}}\quad\text{and}\\
        &A_\text{QS} = \frac{F_0}{k}\, .
\end{align}

\begin{figure}
    \centering
    \includegraphics[width=\columnwidth]{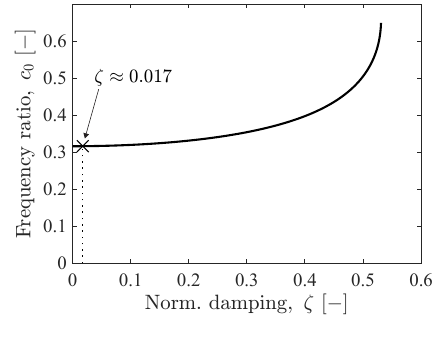}
    \caption{The critical frequency ratio $c_0=f/f_n$ for $\Delta\text{RMS}_\text{th}=0.1$ as a function of nondimensional damping $\zeta$ for the canonical viscously damped, harmonically forced second-order system.}
    \label{fig:E_10}
\end{figure}

Applying the criterion defined in \cref{sec:Defining the boundary between quasi-static and dynamic responses} -- cf.~\cref{eq:Delta_RMS}, we find:
\begin{equation}
    \begin{aligned}
        \Delta \text{RMS} &= \frac{A_\text{dyn}-A_\text{QS}}{A_\text{dyn}} \\ &= 1- \sqrt{(1-c_0^2)^2+(2\zeta c_0)^2}\,.
    \end{aligned}
\end{equation}
Setting $\Delta \text{RMS}_\text{th}=0.1$, we obtain the following biquadratic equation in terms of $c_0$:
\begin{equation}\label{eq:c_0}
    c_0^4 +(4\zeta^2-2)c_0^2+0.19 = 0\,.
\end{equation}
In \cref{fig:E_10}, we plot the solution to \cref{eq:c_0} for the frequency ratio $c_0$ versus the normalized damping $\zeta$. For sufficiently underdamped systems ($\zeta\ll1$), the critical frequency ratio remains approximately constant at $c_0\approx0.31$. 
The reference beam in \cref{table:params} has $\zeta\approx0.017$, well within this $\zeta\ll1$ regime. \cref{table:c_0} tabulates $c_0$ values in the limit $\zeta \rightarrow 0$ for various thresholds $\Delta\text{RMS}_\text{th}$, demonstrating that the critical frequency ratio is constant for a fixed threshold in underdamped systems. These results rationalize the linear relation $f_d^\star = c f_0$ observed in the main text (cf.~\cref{fig:Fig_6}).

\begin{table}[h]
  \centering
  
  \begin{tabular}{cc}
    \toprule
    $\Delta \text{RMS}_{\text{th}}$ & $c_0$ \\
    \midrule
    0.02 & 0.1414 \\
    0.04 & 0.2000 \\
    0.06 & 0.2449 \\
    0.08 & 0.2828 \\
    0.10 & 0.3162 \\
    \bottomrule
  \end{tabular}
  
  \vspace{0.5ex}
  \caption{Critical frequency ratio $c_0$ for a sufficiently underdamped system for the threshold values $\Delta\text{RMS}_\text{th}=0.02,\,0.04,\,0.06,\,0.08,\,0.1$.}\label{table:c_0}
\end{table}

\FloatBarrier


\newpage
\begin{thebibliography}{37}
\expandafter\ifx\csname natexlab\endcsname\relax\def\natexlab#1{#1}\fi
\providecommand{\url}[1]{\texttt{#1}}
\providecommand{\href}[2]{#2}
\providecommand{\path}[1]{#1}
\providecommand{\DOIprefix}{doi:}
\providecommand{\ArXivprefix}{arXiv:}
\providecommand{\URLprefix}{URL: }
\providecommand{\Pubmedprefix}{pmid:}
\providecommand{\doi}[1]{\href{http://dx.doi.org/#1}{\path{#1}}}
\providecommand{\Pubmed}[1]{\href{pmid:#1}{\path{#1}}}
\providecommand{\bibinfo}[2]{#2}
\ifx\xfnm\relax \def\xfnm[#1]{\unskip,\space#1}\fi
\bibitem[{Ahmed et~al.(2022)Ahmed, El~Damatty, Dai, Ibrahim and Lu}]{ahmed2022parametric}
\bibinfo{author}{Ahmed, M.R.}, \bibinfo{author}{El~Damatty, A.A.}, \bibinfo{author}{Dai, K.}, \bibinfo{author}{Ibrahim, A.}, \bibinfo{author}{Lu, W.}, \bibinfo{year}{2022}.
\newblock \bibinfo{title}{Parametric study of the quasi-static response of wind turbines in downburst conditions using a numerical model}.
\newblock \bibinfo{journal}{Engineering Structures} \bibinfo{volume}{250}, \bibinfo{pages}{113440}.
\bibitem[{Apiwattanalunggarn et~al.(2003)Apiwattanalunggarn, Shaw, Pierre and Jiang}]{apiwattanalunggarn2003finite}
\bibinfo{author}{Apiwattanalunggarn, P.}, \bibinfo{author}{Shaw, S.W.}, \bibinfo{author}{Pierre, C.}, \bibinfo{author}{Jiang, D.}, \bibinfo{year}{2003}.
\newblock \bibinfo{title}{Finite-element-based nonlinear modal reduction of a rotating beam with large-amplitude motion}.
\newblock \bibinfo{journal}{Journal of Vibration and Control} \bibinfo{volume}{9}, \bibinfo{pages}{235--263}.
\bibitem[{Baek et~al.(2021)Baek, Johanns, Sano, Grandgeorge and Reis}]{baek2021finite}
\bibinfo{author}{Baek, C.}, \bibinfo{author}{Johanns, P.}, \bibinfo{author}{Sano, T.G.}, \bibinfo{author}{Grandgeorge, P.}, \bibinfo{author}{Reis, P.M.}, \bibinfo{year}{2021}.
\newblock \bibinfo{title}{Finite element modeling of tight elastic knots}.
\newblock \bibinfo{journal}{Journal of Applied Mechanics} \bibinfo{volume}{88}, \bibinfo{pages}{024501}.
\bibitem[{Bazoune(2005)}]{bazoune2005relationship}
\bibinfo{author}{Bazoune, A.}, \bibinfo{year}{2005}.
\newblock \bibinfo{title}{Relationship between softening and stiffening effects in terms of southwell coefficients}.
\newblock \bibinfo{journal}{Journal of sound and vibration} \bibinfo{volume}{287}, \bibinfo{pages}{1027--1030}.
\bibitem[{Behzad and Bastami(2004)}]{behzad2004effect}
\bibinfo{author}{Behzad, M.}, \bibinfo{author}{Bastami, A.}, \bibinfo{year}{2004}.
\newblock \bibinfo{title}{Effect of centrifugal force on natural frequency of lateral vibration of rotating shafts}.
\newblock \bibinfo{journal}{Journal of sound and vibration} \bibinfo{volume}{274}, \bibinfo{pages}{985--995}.
\bibitem[{Burton et~al.(2011)Burton, Jenkins, Sharpe and Bossanyi}]{burton2011wind}
\bibinfo{author}{Burton, T.}, \bibinfo{author}{Jenkins, N.}, \bibinfo{author}{Sharpe, D.}, \bibinfo{author}{Bossanyi, E.}, \bibinfo{year}{2011}.
\newblock \bibinfo{title}{Wind energy handbook}.
\newblock \bibinfo{publisher}{John Wiley \& Sons}.
\bibitem[{Chopra(2012)}]{chopra2012dynamics}
\bibinfo{author}{Chopra, A.}, \bibinfo{year}{2012}.
\newblock \bibinfo{title}{Dynamics of structures: theory and applications to earthquake engineering}.
\newblock Civil Engineering and Engineering Mechanics Series, \bibinfo{publisher}{Prentice Hall}.
\bibitem[{Cilenti et~al.(2024)Cilenti, Cameron and Balachandran}]{CILENTI2024104582}
\bibinfo{author}{Cilenti, L.}, \bibinfo{author}{Cameron, M.}, \bibinfo{author}{Balachandran, B.}, \bibinfo{year}{2024}.
\newblock \bibinfo{title}{Influence of noise on a rotating, softening cantilever beam}.
\newblock \bibinfo{journal}{International Journal of Non-Linear Mechanics} \bibinfo{volume}{159}, \bibinfo{pages}{104582}.
\bibitem[{Clarabut et~al.(2023)Clarabut, Robinson, Poirel and Sarkar}]{clarabut_nonlinear_2023}
\bibinfo{author}{Clarabut, D.}, \bibinfo{author}{Robinson, B.}, \bibinfo{author}{Poirel, D.}, \bibinfo{author}{Sarkar, A.}, \bibinfo{year}{2023}.
\newblock \bibinfo{title}{The nonlinear effects of spinning on the dynamics of a pitching cantilever}.
\newblock \bibinfo{journal}{Journal of Sound and Vibration} , \bibinfo{pages}{117876}.
\bibitem[{Das et~al.(2007)Das, Ray and Pohit}]{das2007free}
\bibinfo{author}{Das, S.}, \bibinfo{author}{Ray, P.}, \bibinfo{author}{Pohit, G.}, \bibinfo{year}{2007}.
\newblock \bibinfo{title}{Free vibration analysis of a rotating beam with nonlinear spring and mass system}.
\newblock \bibinfo{journal}{Journal of Sound and Vibration} \bibinfo{volume}{301}, \bibinfo{pages}{165--188}.
\bibitem[{French(2017)}]{french2017vibrations}
\bibinfo{author}{French, A.P.}, \bibinfo{year}{2017}.
\newblock \bibinfo{title}{Vibrations and waves}.
\newblock \bibinfo{publisher}{CRC press}.
\bibitem[{Grandgeorge et~al.(2021)Grandgeorge, Baek, Singh, Johanns, Sano, Flynn, Maddocks and Reis}]{grandgeorge2021mechanics}
\bibinfo{author}{Grandgeorge, P.}, \bibinfo{author}{Baek, C.}, \bibinfo{author}{Singh, H.}, \bibinfo{author}{Johanns, P.}, \bibinfo{author}{Sano, T.G.}, \bibinfo{author}{Flynn, A.}, \bibinfo{author}{Maddocks, J.H.}, \bibinfo{author}{Reis, P.M.}, \bibinfo{year}{2021}.
\newblock \bibinfo{title}{Mechanics of two filaments in tight orthogonal contact}.
\newblock \bibinfo{journal}{Proceedings of the National Academy of Sciences} \bibinfo{volume}{118}, \bibinfo{pages}{e2021684118}.
\bibitem[{Gutierrez-Prieto et~al.(2024)Gutierrez-Prieto, Gomez and Reis}]{gutierrez-prieto_gyrophilia_2023}
\bibinfo{author}{Gutierrez-Prieto, E.}, \bibinfo{author}{Gomez, M.}, \bibinfo{author}{Reis, P.M.}, \bibinfo{year}{2024}.
\newblock \bibinfo{title}{Harnessing centrifugal and euler forces for tunable buckling of a rotating elastica}.
\newblock \bibinfo{journal}{Extreme Mechanics Letters} , \bibinfo{pages}{102246}.
\bibitem[{Gutierrez-Prieto et~al.(2025)Gutierrez-Prieto, Meulblok, van Hecke and Reis}]{gutierrez-prieto_dynamic_2025}
\bibinfo{author}{Gutierrez-Prieto, E.}, \bibinfo{author}{Meulblok, C.M.}, \bibinfo{author}{van Hecke, M.}, \bibinfo{author}{Reis, P.M.}, \bibinfo{year}{2025}.
\newblock \bibinfo{title}{Dynamic driving enables independent control of material bits for targeted memory}.
\newblock \bibinfo{journal}{arXiv preprint arXiv:2508.16257} \bibinfo{note}{Submitted}.
\bibitem[{Hoskoti et~al.(2023)Hoskoti, Gupta and Sucheendran}]{hoskoti2023modeling}
\bibinfo{author}{Hoskoti, L.}, \bibinfo{author}{Gupta, S.S.}, \bibinfo{author}{Sucheendran, M.M.}, \bibinfo{year}{2023}.
\newblock \bibinfo{title}{Modeling of geometrical stiffening in a rotating blade—a review}.
\newblock \bibinfo{journal}{Journal of Sound and Vibration} \bibinfo{volume}{548}, \bibinfo{pages}{117526}.
\bibitem[{Inman(2009)}]{inman2009engineering}
\bibinfo{author}{Inman, D.}, \bibinfo{year}{2009}.
\newblock \bibinfo{title}{Engineering Vibration}.
\newblock \bibinfo{publisher}{Pearson Education}.
\bibitem[{Jeffcott(1919)}]{jeffcott1919xxvii}
\bibinfo{author}{Jeffcott, H.H.}, \bibinfo{year}{1919}.
\newblock \bibinfo{title}{The lateral vibration of loaded shafts in the neighbourhood of a whirling speed: the effect of want of balance}.
\newblock \bibinfo{journal}{The London, Edinburgh, and Dublin Philosophical Magazine and Journal of Science} \bibinfo{volume}{37}, \bibinfo{pages}{304--314}.
\bibitem[{Johanns et~al.(2021)Johanns, Grandgeorge, Baek, Sano, Maddocks and Reis}]{johanns2021shapes}
\bibinfo{author}{Johanns, P.}, \bibinfo{author}{Grandgeorge, P.}, \bibinfo{author}{Baek, C.}, \bibinfo{author}{Sano, T.G.}, \bibinfo{author}{Maddocks, J.H.}, \bibinfo{author}{Reis, P.M.}, \bibinfo{year}{2021}.
\newblock \bibinfo{title}{The shapes of physical trefoil knots}.
\newblock \bibinfo{journal}{Extreme Mechanics Letters} \bibinfo{volume}{43}, \bibinfo{pages}{101172}.
\bibitem[{Kim and Chung(2016)}]{kim_nonlinear_2016}
\bibinfo{author}{Kim, H.}, \bibinfo{author}{Chung, J.}, \bibinfo{year}{2016}.
\newblock \bibinfo{title}{Nonlinear modeling for dynamic analysis of a rotating cantilever beam}.
\newblock \bibinfo{journal}{Nonlinear Dynamics} \bibinfo{volume}{86}, \bibinfo{pages}{1981--2002}.
\bibitem[{Lacarbonara et~al.(2012)Lacarbonara, Arvin and Bakhtiari-Nejad}]{lacarbonara2012geometrically}
\bibinfo{author}{Lacarbonara, W.}, \bibinfo{author}{Arvin, H.}, \bibinfo{author}{Bakhtiari-Nejad, F.}, \bibinfo{year}{2012}.
\newblock \bibinfo{title}{A geometrically exact approach to the overall dynamics of elastic rotating blades—part 1: linear modal properties}.
\newblock \bibinfo{journal}{Nonlinear Dynamics} \bibinfo{volume}{70}, \bibinfo{pages}{659--675}.
\bibitem[{Lo and Renbarger(1952)}]{lo1952bending}
\bibinfo{author}{Lo, H.}, \bibinfo{author}{Renbarger, J.}, \bibinfo{year}{1952}.
\newblock \bibinfo{title}{Bending vibrations of a rotating beam, first us nat}.
\newblock \bibinfo{journal}{Congr. Appl. Mech., Am. Soc. Mech. Eng} , \bibinfo{pages}{75}.
\bibitem[{Meirovitch(1967)}]{meirovitch1967analytical}
\bibinfo{author}{Meirovitch, L.}, \bibinfo{year}{1967}.
\newblock \bibinfo{title}{Analytical methods in vibrations}.
\newblock \bibinfo{publisher}{Macmillan}.
\bibitem[{Meriam et~al.(2020)Meriam, Kraige and Bolton}]{meriam2020engineering}
\bibinfo{author}{Meriam, J.L.}, \bibinfo{author}{Kraige, L.G.}, \bibinfo{author}{Bolton, J.N.}, \bibinfo{year}{2020}.
\newblock \bibinfo{title}{Engineering mechanics: dynamics}.
\newblock \bibinfo{publisher}{John Wiley \& Sons}.
\bibitem[{Piersol and Paez(2010)}]{piersol2010harris}
\bibinfo{author}{Piersol, A.G.}, \bibinfo{author}{Paez, T.L.}, \bibinfo{year}{2010}.
\newblock \bibinfo{title}{Harris' Shock and Vibration Handbook}.
\newblock \bibinfo{edition}{6th} ed., \bibinfo{publisher}{McGraw-Hill}, \bibinfo{address}{New York, NY}.
\bibitem[{Rajagopal and Mandal(2019)}]{rajagopal2019quasi}
\bibinfo{author}{Rajagopal, A.}, \bibinfo{author}{Mandal, D.K.}, \bibinfo{year}{2019}.
\newblock \bibinfo{title}{A quasi-static solution for the maneuvering analysis of vehicles with rotors}, in: \bibinfo{booktitle}{AIAA Scitech 2019 Forum}, p. \bibinfo{pages}{0863}.
\bibitem[{Rankine(1869)}]{rankine1869centrifugal}
\bibinfo{author}{Rankine, W.M.}, \bibinfo{year}{1869}.
\newblock \bibinfo{title}{On the centrifugal force of rotating shafts}.
\newblock \bibinfo{journal}{Van Nostrand's Eclectic Engineering Magazine (1869-1879)} \bibinfo{volume}{1}, \bibinfo{pages}{598}.
\bibitem[{Rao(2004)}]{rao2004mechanical}
\bibinfo{author}{Rao, S.}, \bibinfo{year}{2004}.
\newblock \bibinfo{title}{Mechanical Vibrations}.
\newblock \bibinfo{publisher}{Pearson Prentice Hall}.
\bibitem[{Shigley et~al.(1985)Shigley, Saunders and Mitchell}]{shigley1985mechanical}
\bibinfo{author}{Shigley, J.}, \bibinfo{author}{Saunders, H.}, \bibinfo{author}{Mitchell, L.}, \bibinfo{year}{1985}.
\newblock \bibinfo{title}{Mechanical Engineering Design}.
\newblock \bibinfo{publisher}{The American Society of Mechanical Engineers (ASME)}.
\bibitem[{Siddiqui et~al.(2017)Siddiqui, Rasheed, Kvamsdal and Tabib}]{siddiqui2017quasi}
\bibinfo{author}{Siddiqui, M.S.}, \bibinfo{author}{Rasheed, A.}, \bibinfo{author}{Kvamsdal, T.}, \bibinfo{author}{Tabib, M.}, \bibinfo{year}{2017}.
\newblock \bibinfo{title}{Quasi-static \& dynamic numerical modeling of full scale nrel 5mw wind turbine}.
\newblock \bibinfo{journal}{Energy Procedia} \bibinfo{volume}{137}, \bibinfo{pages}{460--467}.
\bibitem[{da~Silva and Hodges(1986)}]{da1986nonlinear}
\bibinfo{author}{da~Silva, M.C.}, \bibinfo{author}{Hodges, D.}, \bibinfo{year}{1986}.
\newblock \bibinfo{title}{Nonlinear flexure and torsion of rotating beams, with application to helicopter rotor blades-ii. response and stability results}.
\newblock \bibinfo{journal}{Vertica} \bibinfo{volume}{10}, \bibinfo{pages}{171--186}.
\bibitem[{Southwell and Gough(1921)}]{southwell_free_1921}
\bibinfo{author}{Southwell, R.}, \bibinfo{author}{Gough, F.}, \bibinfo{year}{1921}.
\newblock \bibinfo{title}{The free transverse vibration of airscrew blades}.
\newblock \bibinfo{journal}{British ARC Reports and Memoranda} .
\bibitem[{Strogatz(2018)}]{strogatz2018nonlinear}
\bibinfo{author}{Strogatz, S.H.}, \bibinfo{year}{2018}.
\newblock \bibinfo{title}{Nonlinear dynamics and chaos: with applications to physics, biology, chemistry, and engineering}.
\newblock \bibinfo{publisher}{CRC press}.
\bibitem[{Thomas et~al.(2016)Thomas, S{\'e}n{\'e}chal and De{\"u}}]{thomas2016hardening}
\bibinfo{author}{Thomas, O.}, \bibinfo{author}{S{\'e}n{\'e}chal, A.}, \bibinfo{author}{De{\"u}, J.F.}, \bibinfo{year}{2016}.
\newblock \bibinfo{title}{Hardening/softening behavior and reduced order modeling of nonlinear vibrations of rotating cantilever beams}.
\newblock \bibinfo{journal}{Nonlinear dynamics} \bibinfo{volume}{86}, \bibinfo{pages}{1293--1318}.
\bibitem[{Turhan and Bulut(2009)}]{turhan2009nonlinear}
\bibinfo{author}{Turhan, {\"O}.}, \bibinfo{author}{Bulut, G.}, \bibinfo{year}{2009}.
\newblock \bibinfo{title}{On nonlinear vibrations of a rotating beam}.
\newblock \bibinfo{journal}{Journal of sound and vibration} \bibinfo{volume}{322}, \bibinfo{pages}{314--335}.
\bibitem[{Van~der Walt et~al.(2014)Van~der Walt, Sch{\"o}nberger, Nunez-Iglesias, Boulogne, Warner, Yager, Gouillart and Yu}]{van2014scikit}
\bibinfo{author}{Van~der Walt, S.}, \bibinfo{author}{Sch{\"o}nberger, J.L.}, \bibinfo{author}{Nunez-Iglesias, J.}, \bibinfo{author}{Boulogne, F.}, \bibinfo{author}{Warner, J.D.}, \bibinfo{author}{Yager, N.}, \bibinfo{author}{Gouillart, E.}, \bibinfo{author}{Yu, T.}, \bibinfo{year}{2014}.
\newblock \bibinfo{title}{scikit-image: image processing in python}.
\newblock \bibinfo{journal}{PeerJ} \bibinfo{volume}{2}, \bibinfo{pages}{e453}.
\bibitem[{Wright et~al.(1982)Wright, Smith, Thresher and Wang}]{wright1982vibration}
\bibinfo{author}{Wright, A.}, \bibinfo{author}{Smith, C.}, \bibinfo{author}{Thresher, R.}, \bibinfo{author}{Wang, J.}, \bibinfo{year}{1982}.
\newblock \bibinfo{title}{Vibration modes of centrifugally stiffened beams}.
\newblock \bibinfo{journal}{Journal of Applied Mechanics} \bibinfo{volume}{49}, \bibinfo{pages}{197--202}.
\bibitem[{Yoo and Shin(1998)}]{yoo_vibration_1998}
\bibinfo{author}{Yoo, H.H.}, \bibinfo{author}{Shin, S.H.}, \bibinfo{year}{1998}.
\newblock \bibinfo{title}{Vibration analysis of rotating cantilever beams}.
\newblock \bibinfo{journal}{Journal of Sound and Vibration} \bibinfo{volume}{212}, \bibinfo{pages}{807--828}.

\end{thebibliography}
\end{document}